\documentclass[a4paper,fleqn,usenatbib]{mnras}

\usepackage[T1]{fontenc}
\usepackage{ae,aecompl}

\usepackage{graphicx}	
\usepackage{amsmath}	
\usepackage{amssymb}	
\usepackage{booktabs}
\usepackage{rotating}
\usepackage{caption}

\title[Environmental dependence of X-ray and optical properties of galaxy clusters]{Environmental dependence of X-ray and optical properties of galaxy clusters}

\author[M. Manolopoulou et al.]
{M. Manolopoulou$^{1}$\thanks{E-mail: mmanolop89@gmail.com},
B. Hoyle$^{2}$,
R. G. Mann$^{1}$,
M. Sahl\'en$^{3}$
and S. Nadathur$^{4}$
\\
$^{1}$Institute for Astronomy, University of Edinburgh, Blackford Hill, EH9 3HJ, Edinburgh, UK\\
$^{2}$Universitaets-Sternwarte, Fakultaet fuer Physik, Ludwig-Maximilians Universitaet Muenchen, Scheinerstr. 1,\\ D-81679 Muenchen, Germany\\
$^{3}$Department of Physics and Astronomy, Uppsala University, Box 516, SE- 751 20 Uppsala, Sweden\\
$^{4}$Institute of Cosmology and Gravitation, University of Portsmouth, Burnaby Road, Portsmouth, PO1 3FX, UK
}
\date{Accepted 2020 October 21. Received 2020 October 21; in original form 2020 March 10}
\pubyear{2020}

\begin{document}
\label{firstpage}
\pagerange{\pageref{firstpage}--\pageref{lastpage}}
\maketitle

\begin{abstract}
Galaxy clusters are widely used to constrain cosmological parameters through their properties, such as masses, luminosity and temperature distributions. One should take into account all kind of biases that could affect these analyses in order to obtain reliable constraints. In this work, we study the difference in the properties of clusters residing in different large scale environments, defined by their position within or outside of voids, and the density of their surrounding space.
We use both observational and simulation cluster and void catalogues, i.e. XCS and redMaPPer clusters, BOSS voids, and Magneticum simulations. We devise two different environmental proxies for the clusters and study their redshift, richness, mass, X-ray luminosity and temperature distributions as well as some properties of their galaxy populations. We use the Kolmogorov-Smirnov two-sample test to discover that richer and more massive clusters are more prevalent in overdense regions and outside of voids. We also find that clusters of matched richness and mass in overdense regions and outside voids tend to have higher X-ray luminosities and temperatures. These differences could have important implications for precision cosmology with clusters of galaxies, since cluster mass calibrations can vary with environment.

\end{abstract}


\section{Introduction}

The large-scale structure of the Universe resembles a cosmic web, the density of which is traced by millions of observed galaxies. These galaxies themselves also collapse into galaxy groups and galaxy clusters, which are often located along the walls and filaments of the Cosmic Web. Knots are the intersection of filaments at which point collections of galaxy superclusters are found \citep[for a review see][]{lssreview}. The rest of the space in the Cosmic Web is underdense compared to the walls, filaments and knots, but not devoid of galaxies, and occasionally even galaxy groups and galaxy clusters. These underdense regions are known as voids.

Differences in the properties of the galaxy populations in these different large scale structure environments have been found. \citet{dres80} examined the relationship between local density and galaxy morphology and found indications of increasing elliptical and S0 population and a corresponding decreasing spiral population with increasing density as well as a trend of increasing luminosity of the spheroidal component of galaxies with increasing local density. Additionally, \citet{meneux06} concluded that luminous late-type galaxies are located in more clustered, higher-density regions than are less luminous galaxies. In \citet{ric17}, the authors showed that galaxies which inhabit voids have later galaxy-type morphologies at all stellar masses and that the later-type galaxies appear at smaller distances from the void centre than early-type galaxies. \citet{darvish18} concluded that the molecular gas content and the subsequent star-formation activity of star-forming and starburst galaxies are not affected by their local environment since z$\sim$3.5. In \citet{wang18}, the main sequence of central galaxies and the fraction of star-forming galaxies were found to have no significant dependence on halo mass, while for satellite galaxies the position of the main sequence was found to be almost always lower compared to that of the field and the width is almost always larger. The fraction of star-forming galaxies was seen to decrease with increasing halo mass and this dependence was found stronger towards lower redshift. \citet{hoyle12} found that void galaxies have bluer colours than galaxies in higher density environments with the same magnitude distribution; also an alignment of the disk galaxies angular momenta with the void's radial direction was found in \citet{varela12}. 

Despite the small numbers of galaxies in voids and the typically large void volumes, galaxies still gravitationally attract each other to form groups and clusters of galaxies. The latter are expected to have fewer members and, therefore, smaller sizes and masses and to have undergone fewer mergers in their formation history with respect to the ``field'' clusters, which we define here as groups and clusters not inhabiting voids. As a result of the latter, there would be more relaxed clusters in number within voids than outside voids. \citet{cautun14} used cosmological simulations to find that voids and sheets are devoid of massive clusters. After classifying morphologically the Cosmic Web, \citet{aragoncalvo10} promoted the idea that more massive clusters reside in areas of higher density, while less massive clusters reside in underdense regions. A study of the environmental dependence on the properties of galaxy groups occurred in \citet{poudel16} who found that groups in high-density environments show more efficient galaxy formation and higher abundances of satellite galaxies. \citet{liao18} used hydrodynamical simulations to show that dark matter halos in filaments have higher baryon and stellar fractions than the field counterparts. Any difference in the properties of the clusters in different environmental densities would result in complex and currently unaccounted for selection effects. This would affect studies of galaxy groups and clusters; for example, it could potentially affect the currently adopted scaling relations between cluster observables and cluster masses, the calculation of the cluster power spectrum, which are crucial when using clusters to estimate cosmological parameters \citep[e.g.][]{aem,borgani08}. The differences can be related to the recent conclusions of non-isotropy of the Universe due to spatial variation of the $L$-$T$ cluster relations \citep{migkas20}. Some examples of the reasons why we expect to observe differences in the observed cluster properties as a function of local environment are possibly different merger rates of galaxies within the clusters, gravitational screening mechanisms which modify the force of gravity \citep[e.g.][]{spiegel99} and changes to the cluster formation model \citep[e.g.][]{kravtsov12}.

In this study, we search for differences in the X-ray and optical properties of galaxy clusters as a function of their environment using two methods. In the first we construct void catalogs from the galaxy positions and characterize clusters using this geometrical criterion, i.e. whether they reside inside voids or not. Secondly, we study the differences in cluster properties as a function of local density as directly estimated from the galaxy positions, without using void catalogues. We compare the redshift, richness, X-ray luminosity and temperature distributions of the clusters as well as the BCG and CMR fit properties of their galaxy populations. Some of these properties are widely used to infer the cluster mass: more massive clusters tend to have more galaxy members (i.e., are richer), and have higher temperatures and luminosities in their cores. Therefore, if differences were to be found in these properties between clusters in different large-scale environments, that would imply that an environmental bias correction should be introduced when inferring the cluster mass. Failure to do so could lead to systematically incorrect mass estimates. Since the cluster mass function is very steep, a relatively small systematic bias in the mass estimate could have a large impact on the expected cluster number density \citep{xcscosm}.

To model the large-scale environment we use a common set of voids derived from the Baryon Oscillation Spectroscopic Survey \citep[BOSS,][]{Dawson:2013} spectroscopic galaxy catalogues, and both X-ray \citep[XCS DR2--SDSS,][]{manolopoulou18a} and optically selected \citep[redMaPPer SDSS,][]{rmsdss} cluster samples, which are presented in Section \ref{data}, together with a larger set of catalogues from Magneticum simulation data. In Section \ref{clusterenv}, we identify clusters within and outside voids and calculate the density of their environment. We describe our method of matching samples of different environments and comparing their properties. We then compare the cluster distributions of redshift, richness, mass, luminosity and temperature, brightest cluster galaxy (BCG) and colour-magnitude relation (CMR) fitting parameters within different local environments in Section \ref{resultsvoids}. We create the mass functions of clusters within and outside voids and in overdense and underdense regions. We seek possible differences that would need to be accounted for when doing cosmological analyses using those cluster properties. We study the dependence of the cluster sample size on the results in Section \ref{samplesize}. In Section \ref{discussion}, we present our main results and discuss the effect of the richness estimators in the difference of richness between clusters identified to reside within voids and not. We suggest future prospects of this project and, finally, conclude.

\section{The catalogues}
\label{data}

\subsection{Observational data}

\subsubsection{Void catalogues}

We use a catalogue of voids from the BOSS Data Release 12 galaxy catalogues \citep{Alam-DR11&12:2015}, obtained using the \texttt{REVOLVER} void-finding algorithm \citep{Nadathur:2019c}.\footnote{Available from \url{https://github.com/seshnadathur/Revolver}} \texttt{REVOLVER} is derived from the earlier \texttt{ZOBOV} code \citep{Neyrinck:2008}. The algorithm reconstructs an estimate of the continuous galaxy density field from the discrete tracer distribution using Voronoi tessellations, and then identifies voids as corresponding to minima of this density field, with neighbouring voids delineated based on a watershed algorithm that makes no prior assumption about void shapes. \texttt{REVOLVER} accounts for the complex survey geometry using boundary buffer particles during tessellation, and includes additional corrections for the survey selection function and angular completeness using a weighting scheme described in detail in \cite{nadsdssvoids, nadbossvoids, Nadathur:2019c}. 

Void catalogues for an earlier BOSS data release (DR11) were presented in \citet{nadbossvoids}. The DR12 versions of these catalogues used here have also previously been used in other studies, \citep[e.g.,][]{Nadathur:2016b, Kovacs:2018, Nadathur:2019c, Raghunathan:2019}. The BOSS data consists of two distinct galaxy samples, LOWZ and CMASS, characterized by changes in the targeting, redshift range and sky coverage \citep{Reid-DR12:2016}, and we apply \texttt{REVOLVER} separately to each. In order to achieve completeness of the void catalogues and avoid biasing our numbers of clusters found inside and outside voids, we apply further redshift cuts to obtain 2,968 voids in the redshift range of $0.16\leq z\leq 0.41$ from LOWZ, and 7,057 voids in $0.45\leq z\leq 0.67$ from CMASS. In total, these voids cover $\sim80\%$ of the total available survey volume \citep{nadbossvoids}.

\begin{figure}
	\centering
	\includegraphics[width=\columnwidth]{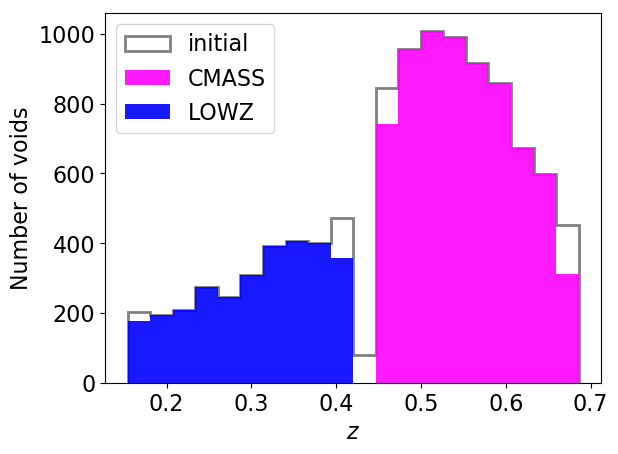} 
    \includegraphics[width=\columnwidth]{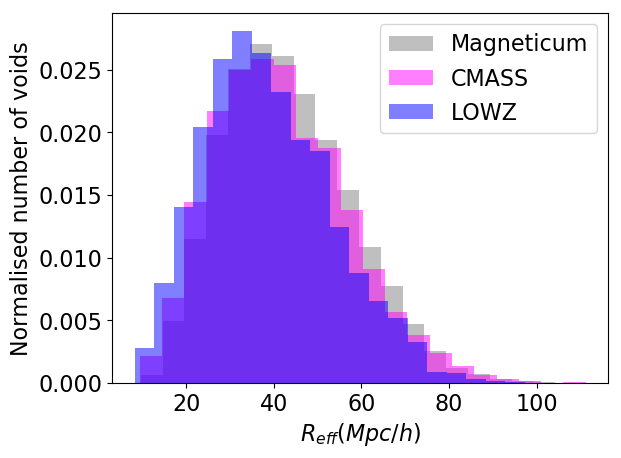}
    \caption{\textit{Top panel:} In grey, the initial redshift distribution of BOSS void catalogues as taken from \citet{nadbossvoids}. The blue and magenta histograms are the redshift-cut distributions of LOWZ and CMASS void catalogues we study respectively. \textit{Bottom panel:} The effective radius distribution of the voids in the two BOSS catalogues (LOWZ in blue and CMASS in magenta) and in Magneticum voids (grey). The lower galaxy number density in CMASS data results in a void size distribution shifted towards larger effective sizes than for LOWZ.}
    \label{fig:voids}
\end{figure}

In the top panel of Figure~\ref{fig:voids} we present the initial BOSS CMASS and LOWZ redshift distributions in grey and the ones after the redshift cuts are applied, in blue and magenta respectively. In the bottom panel of Figure~\ref{fig:voids} we show the normalised distribution of the void effective radius in the two catalogues. CMASS voids have slightly larger sizes than LOWZ voids due to the lower mean galaxy number density, which reduces the spatial resolution of the voidfinder.

The void catalogues contain information about the void centre coordinates, the void effective radius (defined as the radius of a sphere of equivalent volume) and the minimum density within the voids. \texttt{REVOLVER} voids have peculiar 3-dimensional shapes (see \citealt{nadbossvoids} for an example illustration), making a representation of their shape by a sphere unrealistic. We instead model each void in the catalogue as an ellipsoid, and extract information on the lengths of the three ellipsoidal axes and their orientation with respect to the line of sight direction.

\subsubsection{Cluster catalogues}
\label{gmphor}

We use a variety of cluster catalogues in order to explore both X-ray and optical properties of clusters in different environments. We use \textit{(i)} the XMM Cluster Survey (XCS) Data Release 2--Sloan Digital Sky Survey catalogue \citep[XCS DR2--SDSS,][]{manolopoulou18a} to compare cluster X-ray luminosities and temperatures, \textit{(ii)} the GMPhoRCC cluster catalogue, an X-ray selected cluster catalogue from XCS DR2 with optical properties extracted with the Gaussian Mixture full Photometric Red sequence Cluster Characteriser \citep[GMPhoRCC,][]{gmphorcc} and \textit{(iii)} the redMaPPer SDSS DR8 catalogue \citep{rmsdss} to take advantage of the large numbers of galaxy clusters with associated optical properties. We also use the \citet{fino20} redMaPPer cluster subsample which contains clusters X-ray properties as an additional X-ray catalogue. The variation of the catalogue size enables us to study the effect of the sample size in our results.

The XCS DR2--SDSS catalogue is the XCS \citep{xcs1} second data release of X-ray selected galaxy clusters within the SDSS area. The X-ray observations are collected from the XMM public archive and include all areas suitable for cluster searching as described in \citet{dr1xray}. These result in 10,742 observations, each associated with an object ID, across the whole sky. The XCS analysis pipeline is described in detail in Section 2.3 of \citep{gilesbermeo}, which creates an XCS master source list with associated classification of the source apparent size. The extended and Point Spread Function (PSF)-sized sources with number of X-ray soft band (0.5-2 keV) counts higher than 200 are subsequently optically confirmed as galaxy clusters using SDSS DR13 multi-band imaging and a Cluster-Zoo project\footnote{\url{www.zooniverse.org} was used to host our Cluster-Zoo project.}. Within the Cluster-Zoo project, members of the XCS collaboration individually eyeballed and confirmed or rejected the XMM sources as galaxy clusters. This process resulted into an X-ray selected, optically confirmed XCS galaxy cluster catalogue within the SDSS area, that contains 832
galaxy clusters in total. Section 2.5 of \citep{gilesbermeo} describes in detail the measurement of the X-ray properties, including the X-ray bolometric temperature and luminosity (core-included) in the $0.01-50$ keV range of the clusters, which were measured within $R_{500}$, the cluster radius where the density is 500 times higher than the critical density of the Universe. The cluster redshifts were measured using a variety of methods, giving priority to spectroscopic redshift when available. The spectroscopic redshifts were taken from the literature \citep{rmsdss} or were measured using the method of \citet{act2} with data from SDSS DR13 \citep{sdssdr13}, VIPERS PDR2 \citep{vipers} and DEEP2 \citep{deep2}. In the cases where we could not obtain spectroscopic redshift measurements, we measured photometric redshifts using primarily the GMPhoRCC algorithm \citep{gmphorcc} and the zCluster algorithm in the rest of the cases \citep{act2}. The final sample of 832
clusters spans a range of redshifts from $0.025$ to $0.7475$, with median of $0.286$

For a larger number of clusters and the availability of optical properties, we use an extension of the XCS DR2--SDSS catalogue, the GMPhoRCC catalogue. This contains 1,340 clusters with good quality GMPhoRCC flag, associated with optical properties such as red sequence redshift, richness, CMR fitting properties that are calculated with GMPhoRCC \citep{gmphorcc} and X-ray luminosity and temperature calculated as for XCS DR2--SDSS clusters. These have not been optically confirmed through a cluster Zoo like the XCS DR2--SDSS sample or have the X-ray soft band counts threshold, a fact that increases the sample size in the same SDSS area compared to the XCS DR2--SDSS catalogue, but the lower quality of redshifts can contaminate it by including spurious X-ray cluster detections. However, X-ray detected clusters with good quality GMPhoRCC flag coincide with a galaxy overdensity on SDSS catalogue, a fact that optically confirms that they are clusters. 
This catalogue offers a wealth of optical properties to study: red sequence redshift, spectroscopic redshift, where available (coming from the galaxy members with available spectroscopic redshifts; these are usually 1-2 galaxies, but can be up to 5), red sequence colour, CMR width, CMR gradient, CMR intercept, richness within $R_{200}$, BCG distance from the cluster centre (in arcminutes) and finally X-ray temperature within $R_{500}$ and X-ray bolometric luminosity within $R_{500}$ calculated using the same pipelines as the XCS DR2--SDSS catalogue.

\begin{figure}
\centering
	\includegraphics[width=\columnwidth]{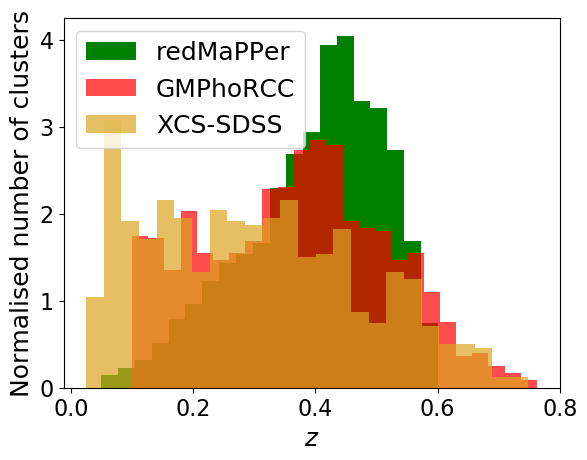}
    \caption{The normalised redshift distributions of the three cluster catalogues. In green, the redMaPPer clusters, in yellow, the XCS DR2--SDSS clusters and in red, the GMPhoRCC clusters.}
    \label{fig:clusterreds}
\end{figure}

The redMaPPer cluster catalogue is an optical catalogue which contains 396,047 galaxy clusters in the SDSS DR8 footprint created with the redMaPPer red sequence cluster finder \citep{rmsdss}. The catalogue contains the cluster redshift $z_{\lambda}$, richness $\lambda$, integrated luminosity in the $i$-band and BCG information (spectroscopic redshift, $i$-band magnitude and i-band luminosity). We use all six available properties in our analysis. As shown in \citep{rmsdss}, the catalogue completeness is richness dependent, therefore, apart from using the full catalogue, we also use the catalogue with extra cuts on the richness and/or redshift:
\begin{itemize}
\item \textit{redMaPPer 1}: $z\geq0.3$ and $\lambda>20$,
\item \textit{redMaPPer 2}: full $z$ range and $\lambda>20$ and
\item \textit{redMaPPer 3}: full $z$ range and $\lambda>30$.
\end{itemize} In Figure~\ref{fig:clusterreds} all three redshift distributions of the cluster catalogues described above are presented.

Additionally, we use a subsample of the redMaPPer cluster catalogue \citep{fino20} which contains the redMaPPer redshift and richeness estimation and also the X-ray luminosity in the rest-frame 0.1-2.4$keV$ and temperature in $keV$. This catalogue contains 10,383 clusters and is used as an additional test for the environmental dependencies on X-ray cluster properties.

\subsection{Simulation data}
\label{magn}

In addition to the observational data, we use simulations to study the properties of large numbers of galaxy clusters inside and outside voids, without the effect of possible detection bias in our cluster and void populations. The Magneticum simulations \citep{Hirschmann14} are large-scale smoothed-particle hydrodynamic (SPH) simulations. 
They are based on the WMAP7 cosmology \citep{komatsu10} and include a variety of physical processes, such as cooling, star formation and stellar winds, chemical enrichment, Active Galactic Nuclei (AGN) feedback and magnetic fields\footnote{For more information refer to \url{http://magneticum.org/}}. For this study, we use the redshift $z=0.14$ snapshot of the Box0 simulation, which has box side 2688 $Mpc$ and $2\times 4536^3$ particles. This redshift is close to but somewhat smaller than that of the BOSS LOWZ data.

The cluster catalogue is created using a friends-of-friends algorithm with a linking length of 0.16 \citep{davis85} that links only the dark matter particles. For each halo, the SUBFIND algorithm \citep{springel01,dolag09} is run in parallel to compute the mass $M$ of the cluster particles within the region where the density is 500 times the critical density of the Universe \citep{magncl}. The centre of each cluster is assigned as its deepest gravitational potential position. The cluster temperature is the mean, mass-weighted temperature within $R_{500}$ and the X-ray luminosity is calculated from the emissivity of every particle in the simulation following \citet{Bartenlann96}. For each cluster in the catalogue we have the mass $M$ within $R_{500}$, the temperature $T_{\rm X}$, and the bolometric X-Ray luminosity within $R_{500}$, $L_{\rm X}$. The catalogue covers a large range of masses, from $10^{11} h^{-1}M_\odot$ to $10^{15} h^{-1}M_\odot$; however, we only use clusters with $M>10^{14} h^{-1}M_\odot$, where the extracted X-ray luminosities and temperatures are reliable (K. Dolag, private communication); this mass cut has also been used for Magneticum clusters in \citet{magncl}. This is a catalogue of $\sim 105,000$ simulated clusters, a much larger X-ray cluster sample than XCS DR2--SDSS and GMPhoRCC, ideal to study differences of X-ray properties of clusters.

To create a void catalogue in the Magneticum simulation data, we first applied a simple galaxy magnitude cut to obtain a sample of simulated galaxies approximately matching the mean number density of the BOSS LOWZ sample, and then used \texttt{REVOLVER} in the same way as for the galaxy data. It should be noted that the simple magnitude cut used means that the simulation galaxy sample does not exactly match the clustering properties of LOWZ galaxies, which also has an effect on the resultant void properties \citep{Nadathur:2015c}. In particular, the bottom panel of Figure~\ref{fig:voids} shows that the Magneticum void sample is slightly shifted towards larger void sizes than LOWZ; however the difference is small and we will neglect it. In total, we obtained 40,000 voids in the simulation, which allows a statistically large sample to compare to the results from BOSS data.

When dealing with Magneticum voids, we bypass the time-consuming determination of the ellipsoidal axes for each void and instead approximate them as spheres. Knowing that this approximation is inaccurate, in Section \ref{clusterenv} we introduce cuts designed to tackle this issue.

\section{Clusters in different environments}
\label{clusterenv}

\subsection{Geometrical sample selection}
Having a variety of cluster and void catalogues we can now begin to study the cluster properties as a function of their environment. In order to determine the environment in which the clusters reside, we use as a probe the location of the cluster within the large scale structure, i.e. whether a cluster is within a void or not.

We use the void catalogues described above to distinguish between clusters outside voids and those within voids. For each cluster and void pair, we determine the distance between the centres of the objects in units of the ellipsoidal axes (for BOSS voids) or the effective spherical radius (for Magneticum voids) and compare it with the distance from the void centre to the nearest boundary of the void. If the cluster-void distance is smaller than the boundary distance, the cluster is assumed to be within the void.

However, for the case of complicated void geometries, even when the above condition is satisfied, the whole of the cluster may not truly lie within the void. In addition, clusters that are close to void boundaries are likely to be part of overdensities within filaments and walls at the void edge and not lie in true underdense regions. To be conservative, we therefore consider the following three cases:
\begin{enumerate}
\item the clusters residing in the inner $70\%$ of the ellipsoidal/spheroid void radius (IV7 category),
\item the clusters residing in the inner $50\%$ of the ellipsoidal/spheroid void radius (IV5 category) and
\item the clusters outside voids (OV category).
\end{enumerate}
We believe that the more conservative the threshold of the distance is the less contamination the sample has from clusters belonging to more overdense regions, hence the IV5 category should contain clusters that are well within the realistic 3-dimensional voids.

\subsection{Density-based sample selection}
In order to remove any concern about the effect of the irregular shapes of the voids on the clusters' assigned environment based on the simple geometrical selection that we discussed above, we also estimate the local densities of the clusters' environments directly, rather than using void location as a proxy.

We calculate the galaxy number density within a shell with 10 Mpc inner and 20 Mpc outer radius from the cluster centre using the physical coordinates of either SDSS DR13 photometric galaxy catalogue \citep{sdssdr13} or the Magneticum galaxy catalogue. We choose this specific shell in order to safely exclude the galaxies that belong to the cluster in the local density calculation. When studying the observational catalogues, we introduce a cut on redshift where the SDSS photometry becomes incomplete, at $z=0.5$. With this, we avoid including biases in the density estimation of clusters with $z>0.5$ from areas where the galaxy population is sparser. We then split the clusters in ten density bins, each bin having equal number of clusters. For the comparisons of cluster properties we consider three cluster categories:
\begin{enumerate}
\item the clusters in the lowest density bin, the clusters in the most underdense regions (LD category),
\item the clusters in the highest density bin, the clusters in the most overdense regions (HD1 category) and
\item the clusters in the second highest density bin, the clusters in overdense regions (HD2 category).
\end{enumerate}
We include the HD2 category as an additional check, because the HD1 bin covers a very broad range of density values (see Figure~\ref{fig:void_density}), which may affect comparisons.

\subsection{Comparing properties}
Having assigned clusters to different environments we are ready to compare the distributions of their various properties. To do so, we use two different non-parametric statistical tests that compare continuous distributions without the need of an input comparison distribution, the Kolmogorov-Smirnov k-sample test \citep[KS hereafter,][]{smi} and the Anderson-Darling test \citep[AD hereafter,][]{adtest}. Both tests measure a ``difference'' between the distributions in question and their default distribution and report a p-value which shows the statistical significance of the result. AD test applies larger weights to the tails of the distributions and therefore is more sensitive in that area. The null hypothesis for both samples is that the samples compared are drawn from the same population.

For every cluster property in question, we compare the distribution of the clusters outside voids (OV category) with one of the two clusters in voids distributions (IV7 or IV5 category) as well as the clusters in underdense regions (LD category) with the clusters in overdense regions (HD1 and HD2 categories). Eventually we have four KS test results, one for each of the four comparisons, and four AD test results for the same comparisons. Non-zero KS or AD statistic value and p-value less than 0.05 means that the null hypothesis of the two tests that the distributions come from the same parental distribution can be rejected with 95\% probability. We initially use both tests for our results, but after ensuring they qualitatively provide the same results, we continue using only KS test which is more efficient computationally.

A KS or AD test p-value lower than 0.05 might not necessarily mean that the given property is different between two different cluster distributions; differences could arise randomly from a selection of clusters in the field population. For that reason, we create a verification test, which samples multiple times randomly the field population of clusters and ensures that the differences found using the KS and AD tests are not due to random selection. 

\subsection{Matched samples}
We want to ensure that any differences between the properties of the cluster populations are not due to the difference of other properties that have already shown signs of differences. Therefore, we will always match the redshift and richness or mass distributions of the clusters in different environments, if available, before comparing their luminosity, temperature or other properties, in order to ensure that any difference found in the latter properties is not a product of differences in the redshift or richness/mass distributions. 

For XCS DR2--SDSS we have three properties in hand, the redshift $z$, the bolometric X-ray luminosity $L_{\rm X}$ and the temperature $T_{\rm X}$. We first compare the redshift between the clusters within and out of voids and clusters in underdense and overdense regions. We use the nearest neighbours algorithm on the normalised cluster redshifts with a linking radius of 0.1 to match each cluster within voids (each cluster in overdense region) with a cluster outside voids (a cluster in underdense region). The linking radius or length represents the maximum distance between the normalised redshifts of two clusters that is used by the nearest neighbours algorithm to consider them matched. The pair of cluster samples in comparison have now the same size and are matched in $z$. We then compare the $L_{\rm X}$ and $T_{\rm X}$ of the matched samples. The same procedure is followed for the Magneticum clusters, with the only difference being that instead of initially matching the samples by their redshift, we match them by the cluster mass $M$. The cluster catalogue comes from the same snapshot in redshift, therefore all clusters have the same redshift.

For GMPhoRCC and redMaPPer the process is more complicated. After matching samples by redshift and achieving same sample size, we compare the richness of the $z$-matched samples. We then use the nearest neighbours algorithm and find for each cluster within voids (cluster in overdense region) the closest cluster outside voids (cluster in underdense region) in the two-dimensional space of redshift and richness. We use their normalised values and a linking length of 0.1. The new samples have the same size as the $z$-matched ones, but they are now matched by both $z$ and richness. Using those samples, we then compare the rest of the properties available for each of the cluster catalogues, such as the $L_{\rm X}$ and $T_{\rm X}$.

\section{Results}
\label{resultsvoids}

\subsection{Number of clusters}
\label{vnumbers}

We search for clusters within and outside voids in both the observational and simulation catalogues and report the number of them on the third and fourth columns of Table~\ref{tab:cluster_numbers}. The number of clusters within and outside voids is the same for each catalogue because we match the cluster samples outside voids to the clusters within voids. We also calculate the density of the cluster environment and report the number of clusters in each of the equal-sized bins on the last column of Table~\ref{tab:cluster_numbers}. We can see that the X-ray selected observational catalogues, XCS DR2--SDSS and GMPhoRCC have a smaller number of clusters, while redMaPPer and Magneticum contain thousands of clusters in each category, ideal for providing statistical significance to our results.

\begin{table}
	\centering
	\caption{Clusters residing within voids identified in the BOSS and Magneticum galaxy catalogues. First column is the name of the cluster catalogue. The second and third columns show the number of clusters within the $70\%$ of the void radius and within the $50\%$ of the void radius. The fourth column shows the number of clusters in each of the equally-sized background density bins of clusters.}
	\label{tab:cluster_numbers}
	\begin{tabular}{lrrr}
	\toprule
		 & IV7 & IV5 & density bin\\
	\midrule
	XCS DR2--SDSS  & 24 & 14 & 69\\
    GMPhoRCC  & 67 & 37 & 105\\
    redMaPPer full  & 34,523 & 20,061 & 31,774\\
    redMaPPer 1 & 5,088 & 2,970 & 3,271\\
    redMaPPer 2 & 5,420 & 3,143 & 3,992\\
    redMaPPer 3 & 1,724 & 992 & 1,534\\
    redMaPPer X-ray & 130 & 76 & 240\\
	Magneticum  & 28,125 & 17,672 & 35,312\\
	\bottomrule
	\end{tabular}
\end{table}

\begin{figure*}
	\centering
	XCS DR2--SDSS \hspace{6cm} GMPhoRCC \\
	\includegraphics[width=.34\textwidth]{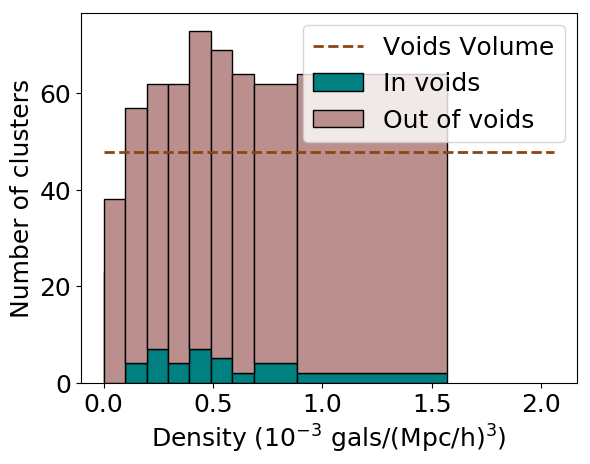}
	\includegraphics[width=.34\textwidth]{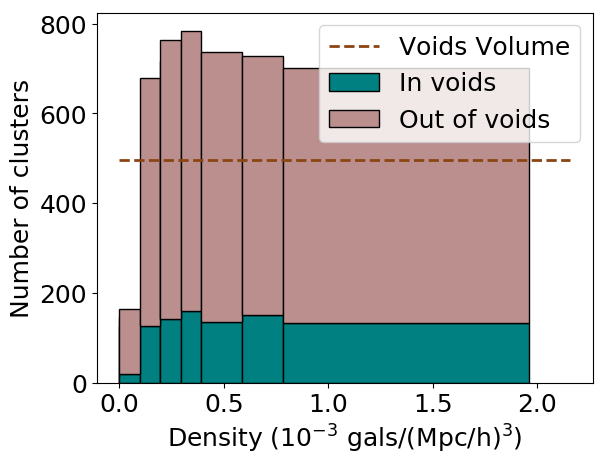}\\
	redMaPPer full \hspace{7cm} Magneticum \\
	\includegraphics[width=.34\textwidth]{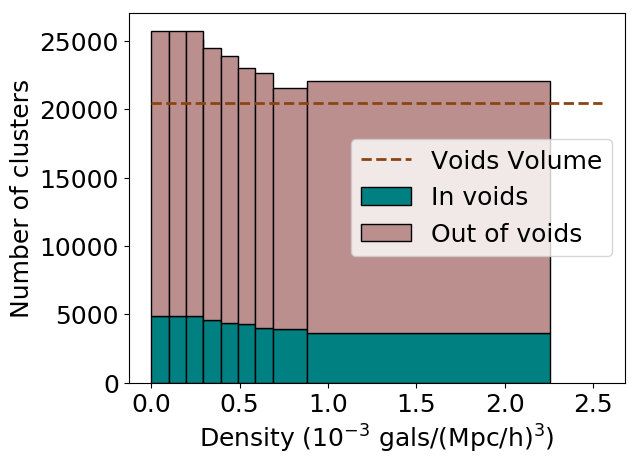}
	\includegraphics[width=.34\textwidth]{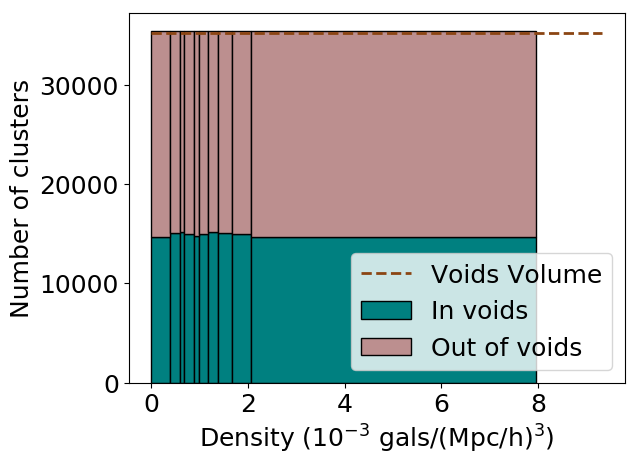}
	\caption{The number of clusters in and out of voids in each density bin for XCS DR2--SDSS (top left), GMPhoRCC (top right), redMaPPer full (bottom left) and Magneticum (bottom right) cluster catalogue. The brown dashed line shows the number of clusters we would expect in voids considering the percentage of the volume survey that are voids.}
	\label{fig:void_density}
\end{figure*}

For each of the cluster catalogues we count the number of clusters of each density bin that are found within and outside voids (Figure~\ref{fig:void_density}). One would expect that clusters with low background density would more likely be found within voids and that clusters with high background density would be more likely found outside clusters, i.e. that more clusters would be found within voids in the low background density bins and vice versa. Looking at Figure~\ref{fig:void_density}, in general, the background cluster density does not seem to relate to whether the cluster resides within a void or not. For the redMaPPer full catalogue, we find that the number of clusters found inside voids is larger at lower background densities, and smaller at high background density. For the other catalogues, the number of clusters in voids is approximately constant across all density bins, showing that splitting clusters within and outside large voids does not necessarily trace the density of their local environment.

The brown dashed horizontal lines on Figure~\ref{fig:void_density} represent the number of clusters we would expect to find within voids considering the percentage of the survey volume contained within voids in each case. The fact that this number is much higher than the number of clusters in voids found is possibly due to the fact that clusters are assumed to be ellipsoidals (for XCS DR2--SDSS, GMPhoRCC and redMaPPer full) or spheroids (for Magneticum) as opposed to their true 3-dimensional shapes. Voids with that approximated shape may overlap with each other and contain less clusters overall compared to the number they would contain if they had their real 3-dimensional shape. This reinforces the decision of having conservative cuts when studying clusters in voids; we have only used clusters within 50\% or 70\% the void radius.

We note here that summing the number of clusters within and outside voids in each density bin in the observational catalogues in Figure~\ref{fig:void_density} does not result to the number of clusters in each density bin that is shown in Table \ref{tab:cluster_numbers}. This is because the clusters within/outside voids have gone through a redshift cut in order to match the redshift range of the BOSS voids catalogue. Figure~\ref{fig:void_density} also confirms the importance of our conservative cuts when studying clusters in voids; we have only used clusters within 50\% or 70\% the void radius.

\subsection{Redshift distributions}
\label{vredshift}

As explained earlier, we begin with comparing the redshift distribution of clusters in the observational catalogues between the clusters within/outside voids and in overdense/underdense regions. For both XCS DR2--SDSS and GMPhoRCC, no differences are found when looking at cluster environment by geometrical criteria, but significant differences that are not associated with random selection were found when comparing clusters by their background density. The lack of differences seen in the former case might be a reflection of the low number statistics of clusters within and outside voids in both catalogues. On the other hand, redMaPPer clusters in different environments present significant differences in their redshift distributions, no matter how the environment is defined and in the full, all three subsamples and the X-ray subsample of the catalogue. In all cases where differences were found, it is seen that more clusters reside within low-density environments in low redshifts, a fact that could be explained by the expansion of the Universe and our use of physical coordinates to calculate the clusters' local density.

\begin{figure*}
	\centering
	XCS DR2--SDSS \hspace{5cm} GMPhoRCC \hspace{5cm} redMaPPer full\\
	\includegraphics[width=.32\textwidth]{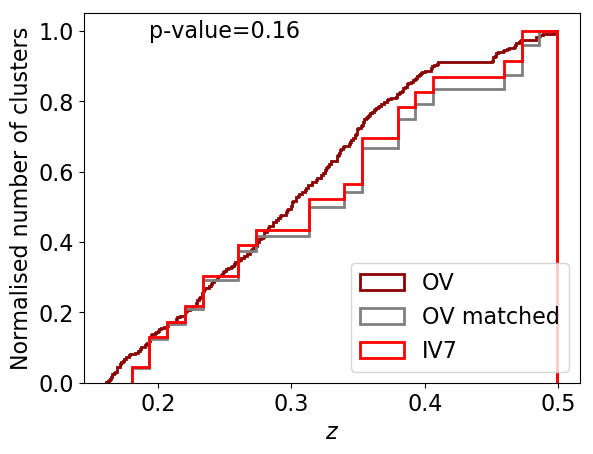}
	\includegraphics[width=.32\textwidth]{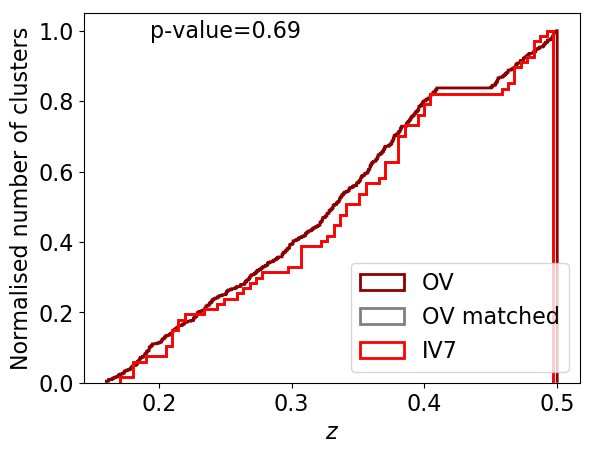}
	\includegraphics[width=.32\textwidth]{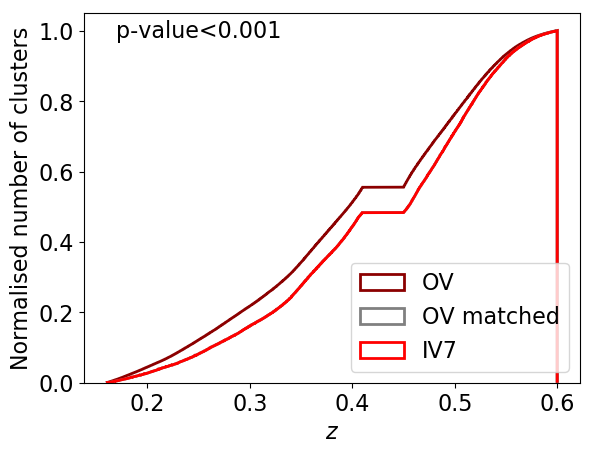}
	\includegraphics[width=.32\textwidth]{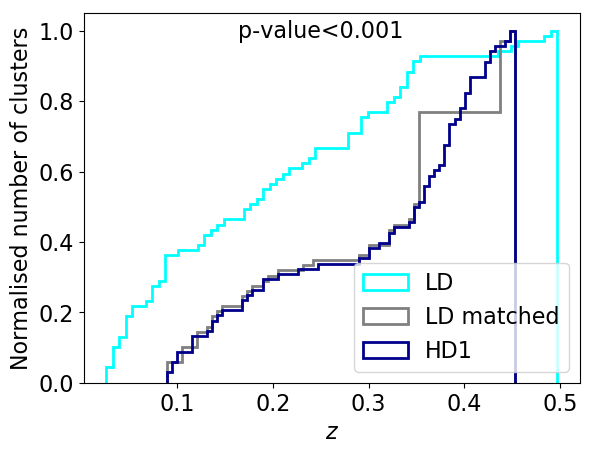}
	\includegraphics[width=.32\textwidth]{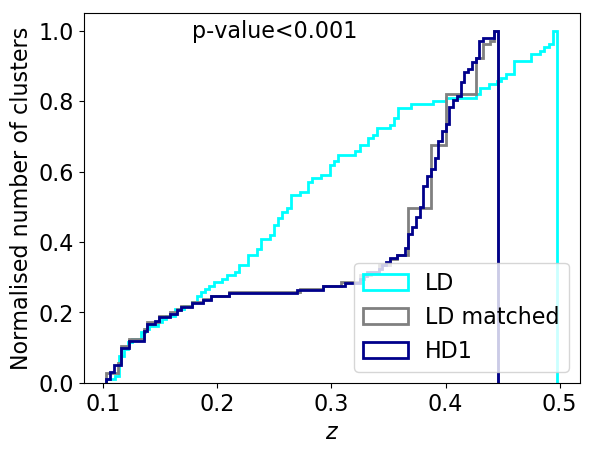}
	\includegraphics[width=.32\textwidth]{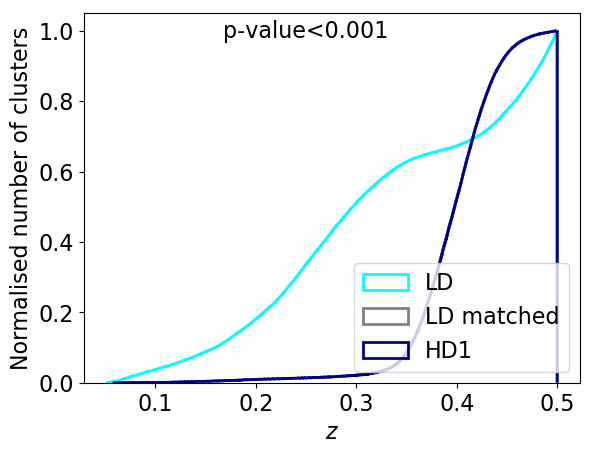}
    \caption{The normalised cumulative redshift distributions of XCS DR2--SDSS (left), GMPhoRCC (middle) and redMaPPer full (right) clusters in IV7 and OV categories (upper panels) and most overdense and most underdense regions (lower panels). Once redshift distributions are compared, the OV and LD categories are matched to the IV7 and HD1 ones respectively and the p-values of their comparisons are shown on the top of the graphs. The redMaPPer matched distributions match the IV7 and HD1 distributions very well and therefore are not distinguishable in the graph.}
    \label{fig:redshift_hists}
\end{figure*}

The different redshift distributions between the cluster samples confirm the need to use matching samples in redshift from now on. Any difference found between other cluster property distributions will not be an effect of the different redshift distributions. The redshift distributions compared, along with the matched ones, are shown in Figure~\ref{fig:redshift_hists}. The distributions on the left panels flatten at $z\sim 0.4$ because of the absence of voids at that redshift range - it is roughly the high-$z$ end of the LOWZ voids and the lowz-$z$ end of the CMASS voids.

\subsection{Mass and richness distributions}
\label{vrichness}

\begin{figure}
	\centering
	Magneticum \\
	\includegraphics[width=.8\columnwidth]{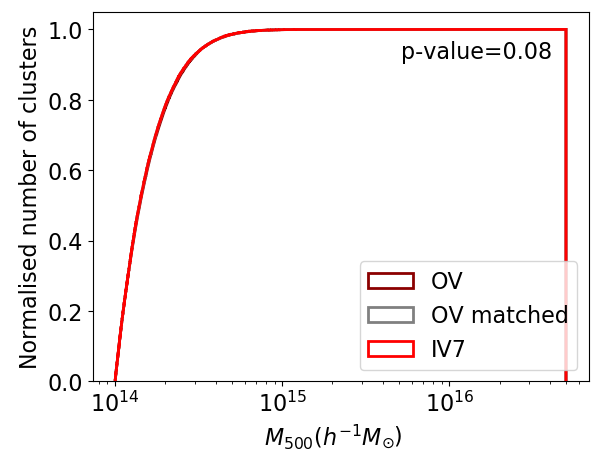}
	\includegraphics[width=.8\columnwidth]{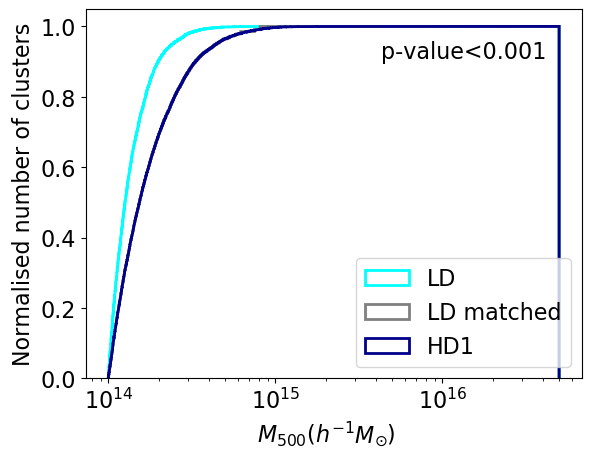}
    \caption{The normalised cumulative mass distributions of Magneticum clusters in IV7 and OV categories (top) and most overdense and most underdense regions (bottom). Once mass distributions are compared, the OV and LD categories are matched to the IV7 and HD1 ones respectively and the p-values of their comparisons are shown on the top of the graphs.}
    \label{fig:mass_hists}
\end{figure}

For Magneticum clusters, we have a single redshift slice, hence the redshift of all clusters is the same, we perform matching on the cluster mass. We find that the definition of the environment here plays a role in the comparisons. The mass distributions of clusters with different background densities are significantly different from each other, while there is no difference found between the clusters within and outside Magneticum voids. This is a hint that the geometrical classification of the cluster environment might be naive, given the irregular shape of voids that is here modeled as spherical and the fact that Magneticum contains a larger amount of voids than the BOSS catalogues. Once again, after comparing the mass distributions, we match the catalogues to have as similar mass distributions as possible before we compare other properties. The distributions compared along with the matched ones are shown in Figure~\ref{fig:mass_hists}. 

For GMPhoRCC and redMaPPer, where richness estimators of the clusters are available, $n_{200}$ and $\lambda$ respectively, we compare those between clusters in different environments with matched redshift distributions. For GMPhoRCC catalogues, significant differences between the richness estimator of the clusters are found between clusters in overdense and underdense regions (HD1-LD $p$-value $<0.001$), but not between clusters within and outside voids (IV7-OV $p$-value $=0.41$). For the redMaPPer catalogue, in both definitions of environment and in all four (sub)catalogues and the X-ray subsample, the richness distributions of clusters present significant differences, with KS $p$-values very close to zero (HD1-LD and IV7-OV $p$-value $<0.001$). After comparing the richness distributions, we match the samples in both redshift and richness for the GMPhoRCC and redMaPPer catalogues, before moving on to compare the rest of the properties.

Those comparisons show clear signs that clusters inside voids and in underdense regions have lower number of galaxy members but similar redshift distribution to clusters outside voids and in overdense regions respectively.

It is worth noting here that the difference found between the richness distributions of clusters in different environments could be an artefact of the algorithm used to calculate the cluster richness. Concerning the redMaPPer catalogue, it has been shown that, during the richness estimation, projection effects depend on both the background galaxy density field and the large cluster-to-cluster fluctuations on the density field. The former only boosts the cluster richness by an unimportant amount and the latter can affect severely 5-15\% of the clusters \citep{rmr}. The richness estimation of a small percentage of redMaPPer clusters might be affected by the large scale structure density field and hence the presence of a cluster in a void or an overdensity. As for the XCS DR2--SDSS and GMPhoRCC catalogues, the richness bias inserted by GMPhoRCC during the richness calculation has also been estimated and it is smaller than the error of the richness calculation, therefore not significant \citep[see calculation in][]{manolopoulouthesis}. 

We can see if the differences found between the different cluster populations can significantly affect cosmological studies that use galaxy clusters by measuring the cluster richness functions for the redMaPPer clusters and the mass functions for the Magneticum clusters. XCS DR2--SDSS and GMPhoRCC catalogues do not have large enough numbers of clusters for this purpose. The richness functions are good approximation of the cluster mass functions since richness has proven to be a good mass proxy for galaxy clusters for redMaPPer SDSS DR8 clusters \citep{richmassrel}. 

\begin{figure*}
    redMaPPer full \hspace{7cm} Magneticum \\
    \includegraphics[width=.34\textwidth]{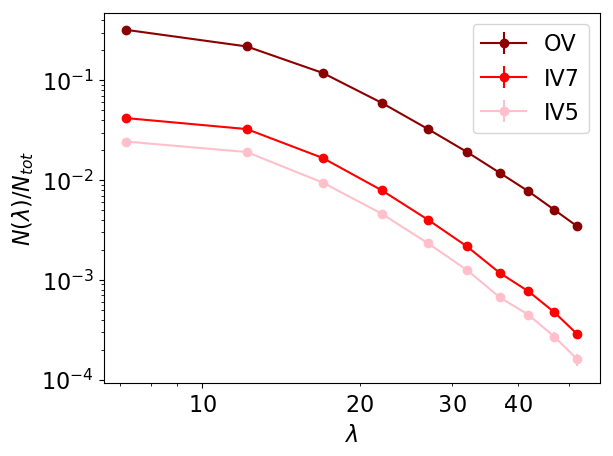}
	\includegraphics[width=.34\textwidth]{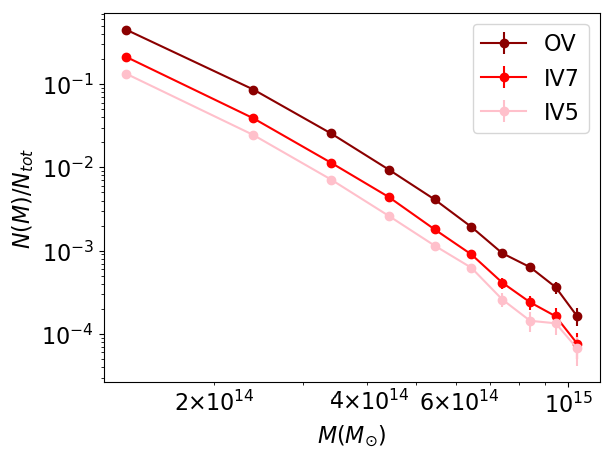}
	\includegraphics[width=.34\textwidth]{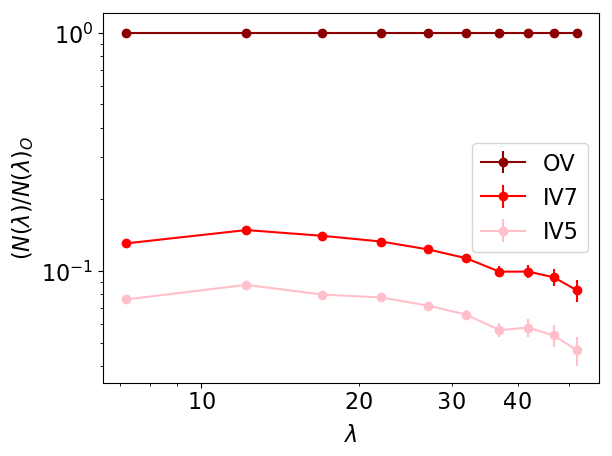}
	\includegraphics[width=.34\textwidth]{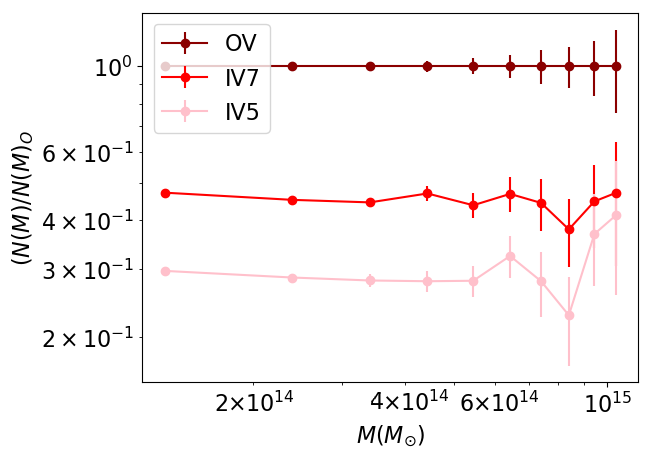}
    \includegraphics[width=.34\textwidth]{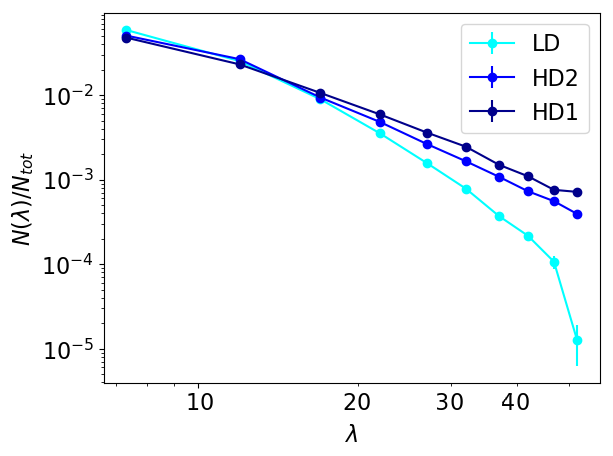}
	\includegraphics[width=.34\textwidth]{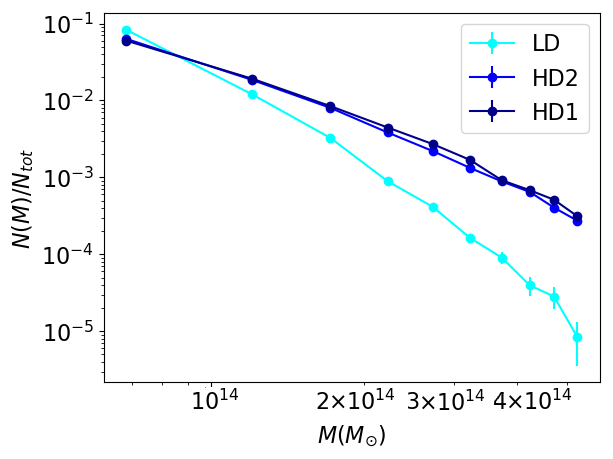}
	\caption{The normalised richness and mass functions of redMaPPer full (left) and Magneticum (right) clusters respectively with respect to the number of all clusters available in OV, IV7 and IV5 categories (top panels), the normalised functions to the number of clusters in OV category (clusters outside voids; middle panels) and the number of all clusters available in LD, HD1 and HD2 categories (bottom panels). It is easier to see the discrepancies of the functions in the high richness/mass end of the top panels in the middle ones. Errors are included in all functions.}
    \label{fig:mass_functions}
\end{figure*}

We construct the richness functions of the redMaPPer clusters and the mass functions of the Magneticum clusters and present them in Figure~\ref{fig:mass_functions}. The top panels show the richness and mass functions of redMaPPer (full) and Magneticum clusters respectively within and outside voids and the bottom ones show the same functions but for clusters in overdense and underdense regions. The middle panels are constructed as the top ones, but the clusters within voids functions are normalised with respect to the clusters outside voids; this is to make comparison between the three functions easier in their high richness/mass end. 

Looking at Fig.~\ref{fig:mass_functions}, especially at the bottom panels, the most massive and rich clusters are found in the most overdense regions, while the same number of low richness/mass clusters appears in overdense and underdense regions. These results are in agreement with the theoretical work based on N-body simulations of the large scale structure in \citet{cautun14}. 
This difference found between clusters within different density environments is a hint that, when selecting clusters by their mass for cosmological studies, a selection bias can enter caused by the environment of the clusters.

\subsection{Luminosity and temperature distributions}
\label{vlt}

\begin{figure}
	\centering
	redMaPPer X-ray \\
	\includegraphics[width=.8\columnwidth]{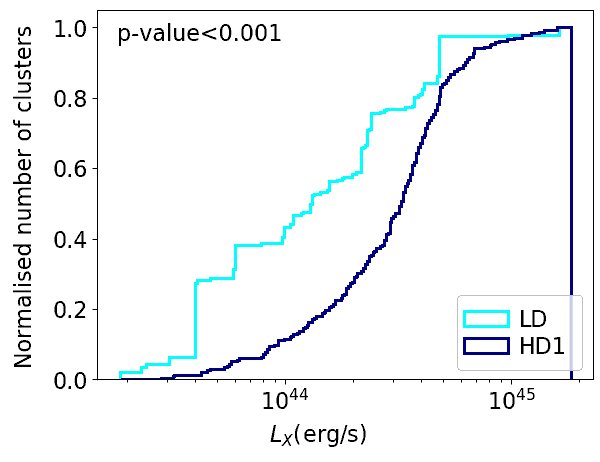}
	\includegraphics[width=.8\columnwidth]{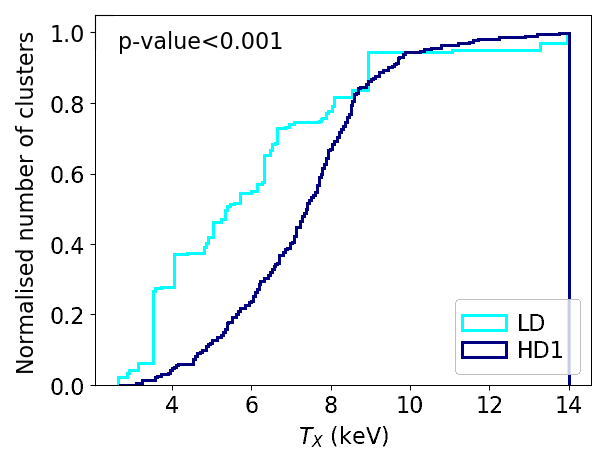}
    \caption{The normalised cumulative X-ray luminosity (top) and temperature (bottom) distributions of redMaPPer X-ray clusters in most underdense (LD) and most underdense (HD1) cluster categories.}
    \label{fig:lt_hists}
\end{figure}

Having matched samples in all of our cluster catalogues, whether by redshift, mass or redshift and richness, we can now study differences in the luminosities and temperatures of the clusters in different large scale environments independently of the matched properties. We remind the reader that for XCS DR2--SDSS, GMPhoRCC, Magneticum and redMaPPer X-ray cluster catalogues we compare the cluster X-ray luminosity and temperature, while for redMaPPer full and its subsamples, we compare the clusters' $i$-band luminosity. For GMPhoRCC catalogue in both definitions of the cluster environment, we find significant differences between the X-ray luminosity of the clusters in overdense and underdense regions and within and outside voids that are confirmed by our random test (HD1-LD and IV7-OV $p$-value $<0.001$). For redMaPPer X-ray,  we find significant differences of the X-ray luminosity distributions between the OV-IV7 samples (KS $p$-value $=0.007$) and not between the OV-IV5 samples (KS $p$-value $=0.104$) - we also find significant differences between the HD1-LD and HD2-LD samples (KS $p$-value $<0.001$ in both cases). The distributions of X-ray luminosity of the HD1 and LD samples are shown in Fig.~\ref{fig:lt_hists}. For XCS DR2--SDSS and Magneticum catalogues we find significant differences between the X-ray luminosity of the clusters in overdense and underdense regions only, confirmed by our random test. For XCS DR2--SDSS, in the HD1-LD case the KS $p$-value is smaller than 0.001 and in OV-IV7 case the KS $p$-value is 0.62. For Magneticum, in the HD1-LD case the KS $p$-value is smaller than 0.001 and in OV-IV7 case the KS $p$-value is 0.55. Voids and underdense regions host higher numbers of low luminosity clusters compared to overdense regions and regions outside voids.  

X-ray temperatures show more consistent between catalogues. When splitting clusters within and outside voids, in all cluster catalogues (XCS DR2--SDSS, GMPhoRCC, redMaPPer X-ray and Magneticum) no significant differences are found in the X-ray temperature distributions of clusters within/outside voids (OV-IV7 KS $p$-value $\sim$ 0.21, 0.41, 0.05 and 0.62 respectively). When splitting clusters according to the density of their environment, significant differences are found in all catalogues in the X-ray temperature distributions of clusters in overdense and underdense regions, confirmed by our random test (all HD1-LD KS $p$-values $<0.001$). In Fig.~\ref{fig:lt_hists} we present the distributions of X-ray temperature of the HD1 and LD samples. High-temperature clusters are more prevalent in overdense regions and low-temperature clusters are more prevalent in underdense regions.

\subsection{Cluster galaxy populations}
\label{optical_properties}

GMPhoRCC and redMaPPer catalogues contain some additional optical properties of the clusters and allow us to study possible differences of those properties in clusters within different environments. GMPhoRCC optical properties contain the CMR fitting properties and the colour of the red sequence of the clusters \citep[][for more information on these properties]{gmphorcc}, while redMaPPer contains the $i$-band cluster luminosity as well as the $i$-band magnitude and luminosity of the BCG of the clusters. 

The CMR intercept, gradient and width distributions of clusters in overdense and underdense regions present significant differences confirmed by the random test (HD1-LD KS $p$-value $\sim$ $1.5\times10^{-3}$, $<0.001$ and $6.7\times10^{-3}$ respectively); the red sequence colour distributions are significantly different between clusters in different environments no matter the environment definition (HD1-LD and IV7-OV KS $p$-value $<0.001$). Those results suggest that clusters in overdense regions tend to have flatter and narrower CMRs and redder $g-r$ colours, meaning that the environment is affecting the properties of the galaxy populations in the clusters as well as of the intracluster gas.
When looking at the BCG properties ($i$-band magnitude and luminosity) and the $i$-band cluster luminosity, those are significantly brighter in the $i$-band in clusters in overdense regions and outside voids compared to clusters in underdense regions and inside voids (all KS $p$-values are smaller than 0.001).

\section{Sample size}
\label{samplesize}

The redMaPPer and Magneticum cluster catalogues offer a large number of clusters, large enough to enable the study of the dependence of the detected differences between two distributions of cluster properties to the number of clusters available. To this end, we take 100 random subsamples of various sizes from each of the two cluster catalogues and compare clusters within/outside voids and in overdense/underdense regions. We present the number of realisations we find significant differences between the cluster samples.

We choose to compare the luminosity distributions ($i$-band for redMaPPer full and X-ray for Magneticum) of clusters within 70\% of the voids radius and outside voids and of clusters in the most overdense and most underdense regions. In Table \ref{tab:sample_size} we show the number of times we find that the luminosity distributions are significantly different when we use random subsamples of the initial catalogues. The sample size seems to affect the two catalogues in different degree. For Magneticum, the difference in the X-ray luminosity distributions of clusters within and outside voids has already disappeared when we take 70\% of the initial Magneticum cluster sample. In the case of comparing overdense and underdense regions, the difference signal is degrading slower and smoother; it is fully observed when taking 70\% of the initial sample and degraded to half when taking 20\% of the initial sample. In the redMaPPer full catalogue, the signal is a lot more persistent compared to the Magneticum catalogue signal and slowly degrades with the sample size when we compare the within voids and outside voids cluster distributions. The signal remains fully detected in all sample sizes when we compare clusters in overdense and underdense regions. The fact that the geometrical environmental proxy is more sensitive than the local density one can be explained by the fact that the samples of clusters within voids are smaller than the similar density ones, as can be seen in Table \ref{tab:cluster_numbers}. The fact that the signal is more sensitive in the Magneticum catalogues compared to the redMaPPer one could be due to the different way of defining the clusters and their properties in the two catalogues.

\begin{table}
\centering
\caption{The number of realisations (out of 100) where the luminosity distributions of redMaPPer and Magneticum clusters in different environments are found significantly different. For the Magneticum catalogue, we take subsamples of 70\%, 50\% and 20\% of the initial catalogue, while for redMaPPer we take subsamples of 50\%, 10\% and 5\% of the initial catalogue.}
\begin{tabular}[c]{cccc}
	\multicolumn{4}{c}{\textbf{Magneticum}} \\
\toprule
	 & 70\% & 50\% & 20\% \\
\midrule
	OV-IV7 comparison & 8 & 3 & 2 \\
	LD-HD1 comparison & 100 & 95 & 52 \\
\bottomrule
	\multicolumn{4}{c}{\textbf{redMaPPer full}} \\
\toprule
	 & 50\% & 10\% & 5\% \\
\midrule
	OV-IV7 comparison & 100 & 71 & 38 \\
	LD-HD1 comparison & 100 & 100 & 100 \\
\bottomrule
\end{tabular}
\label{tab:sample_size}
\end{table}

These results show that the number of clusters available for those comparisons affect the signals observed. The small number of clusters in the XCS DR2--SDSS catalogue are insufficient to give statistically significant results on the difference between the X-ray properties of clusters within and out of voids. As seen in Section \ref{resultsvoids}, there are cases where amongst all the catalogues, XCS DR2--SDSS was the only one that no differences have been observed. A larger X-ray cluster catalogue would be needed for this study, which will be available with eRosita mission \citep{erosita}.

\section{Conclusions}
\label{discussion}

In this work, we studied the difference of the distributions of some main properties of galaxy clusters that reside in various environments, either defined by geometrical criteria or the local density. Our main findings include:

\begin{itemize}
    \item The redshift distributions of clusters within environments of different densities present significant differences - there are more clusters within low-density environments in low redshifts compared to the high-density environments. No significant differences were found for clusters residing within and outside voids. Those results are consistent for XCS DR2--SDSS, GMPhoRCC and redMaPPer catalogues.
    \item The mass distribution of Magneticum clusters and the richness distribution of GMPhoRCC and redMaPPer clusters depend on the environment, with more massive and richer clusters being more prevalent in more overdense regions and outside of voids. 
    \item X-ray luminosity and temperature distributions of clusters seem to differ within different environments across all the catalogues, with clusters with higher luminosity and temperature more likely to appear in overdense regions and outside voids. 
    \item Clusters in overdense regions tend to have flatter and narrower CMRs, with redder $g-r$ colours and more luminous BCGs when matched by redshift, as revealed from the GMPhoRCC and redMaPPer catalogues, so that the environment is affecting the properties of the galaxy populations in the clusters as well as of the intracluster gas.
    \item Local density-defined cluster samples often yield significant results when the geometrically-defined samples do not, suggesting that the former is more physically-motivated way to define samples than the latter.
    \item Possible implications on the cluster mass calibrations in various environments could occur as a result of the above, that would consequently affect cosmological parameters estimated based on those mass calculations.
\end{itemize}

In all cases, the sense of these differences is as expected from our initial intuition about clusters in underdense regions having had quieter merger histories, so that they could accrete less mass through mergers over their lifetime. The results concerning the mass and richness distributions are in agreement with the theoretical work presented in \citet{cautun14}, where underdense regions were found to be devoid of massive galaxy clusters. Similar conclusions were shown in \citet{aragoncalvo10}, where the Cosmic Web was classified using morphological criteria instead of density criteria. 

In \citet{farahi19}, it is shown that clusters with lower X-ray luminosities and younger BCGs have also lower X-ray temperatures, which is in qualitative agreement with our results of difference between the X-ray luminosity and temperature distributions of clusters within overdense and underdense regions. The signal of differences on the X-ray temperature distributions is not present when looking at clusters within and outside voids, while it is detectable for the case of X-ray luminosity distributions, possibly due to the difference in scatter between the two properties (as presented in Table 2 in \citet{farahi19}). The fact that the signal is not present for temperature distributions between clusters within and outside voids could be due to the difficulty of approximating general non-spherical void shapes as ellipses/spheres, as that might also increase the scatter of the properties.

Similar findings to ours were reported by other studies of differences in properties of galaxies in various environments \citep[e.g.][]{hoyle12}. Just like their host clusters, galaxies in clusters in underdense regions seem to be younger, hence their bluer colour, and as a result spirals, as opposed to more old, elliptical galaxies found in clusters outside voids and in overdense regions. The cluster environment seems to affect the evolution of their galaxy members, not only their BCG. These effects could have important implications for precision cosmology with clusters of galaxies, since cluster mass calibrations can vary with environment. Further research with larger observational X-ray cluster catalogues such as from the future eRosita mission \citep{erosita} will be invaluable to further study this effect with more accuracy and give more definitive results.

\section*{Acknowledgements}
We thank the referee for the useful comments that helped to improve the publication. M.M. would like to thank D. Rapetti for the useful discussions and ideas on the cluster mass functions. M.S. was supported by the Olle Engkvist Foundation Project No. 2016/150, the Lundstr\"{o}m--\r{A}man Foundation, and a P.~E.~Fil\'{e}n fellowship (Uppsala University). S.N. was supported by UK Space Agency grant ST/N00180X/1.

\section*{Data availability}
The data underlying this article will be shared on reasonable request to the corresponding author.

\bibliographystyle{mnras}
\bibliography{refs}

\appendix

\bsp	
\label{lastpage}
\end{document}